\documentclass{cernrep}
\usepackage[symbol]{footmisc}

\newcommand{\mttbar}{t\bar{t}}
\newcommand{\mbbbar}{b\bar{b}}
\newcommand{\ttbar}{\ensuremath{t\bar{t}}\xspace}
\newcommand{\bbbar}{\ensuremath{b\bar{b}}\xspace}
\newcommand{\nunubar}{\ensuremath{\nu\bar{\nu}}\xspace}

\usepackage{color}
\definecolor{OliveGreen}{cmyk}{0.64,0,0.95,0.40}
\definecolor{Brown}{cmyk}{0,0.81,1,0.60}
\definecolor{RoyalBlue}{cmyk}{0.71,0.53,0,0.12}
\usepackage[colorlinks=true, linkcolor=OliveGreen, citecolor=RoyalBlue,
urlcolor=Brown]{hyperref}

\begin{document}
\vspace*{2cm}
\centerline{\textbf{\Large{Charting the European Course to the High-Energy Frontier}}}
\vspace{4mm}
U.~Amaldi$^1$,
E.~Aslanides$^2$,
R.~Barate$^{3}$,
C.~Benvenuti$^4$,
P.~Bloch$^{4,5}$,
T.~Camporesi$^{4}$,
A.~David$^{4}$,
D.~Denegri$^{6}$,
M.~Diemoz$^{7}$,
L.~Di~Lella$^{8}$,
G.~Dissertori$^{9}$,
N.~Doble$^4$,
J.~Dumarchez$^{10}$,
J.~Ellis$^{11,12,4}$,
J.~Engelen$^{13}$,
C.~Fabjan$^{14,4}$,
B.~Fuks$^{15}$,
P.~Gavillet$^{4}$,
A.~Hoecker$^4$,
J.~Iliopoulos$^{16}$,
P.~Innocenti$^4$,
W.~Kozanecki$^6$,
P.~Lebrun$^{4}$,
C.~Llewellyn~Smith$^{17}$,
C.~Louren\c{c}o$^4$,
L.~Maiani$^{7}$,
L.~Malgeri$^4$,
M.~Mangano$^{4}$,
F.~Moortgat$^{4}$,
M.~Mulders$^{4}$,
S.~Myers$^{18}$.
F.~Nessi-Tedaldi$^9$,
L.~Pape$^{4}$,
F.~Pauss$^{9}$,
R.~Perin$^4$,
T.~Rodrigo$^{19}$,
G.~Rolandi$^{20}$,
A.~Schopper$^4$,
J.~Schukraft$^{21}$,
M.~Spiro$^{6}$,
L.~Sulak$^{22}$,
T.~Taylor$^{4}$,
D.~Treille$^{4}$,
G.~Tonelli$^{8}$,
G.~Unal$^4$,
F.~Vannucci$^{10}$,
J.~Varela$^{23}$,
T.~Virdee$^{5,4}$,
R.~Voss$^4$,
R.~Wallny$^9$,
H.~Wenninger$^4$,
A.~Zalewski-Bak$^{24}$,
J.~Zinn-Justin$^{6}$\\
~~\\
%
\noindent
$^1$~\small{TERA Foundation, Novara, Italy},
$^2$~\small{Aix Marseille Universit\'e, Marseille, France}, 
$^{3}$~\small{LAPP, Annecy, France}, 
$^{4}$~\small{CERN, Geneva, Switzerland},
$^5$~\small{Imperial College, London, UK},
$^{6}$~\small{IRFU CEA, Saclay, France}, 
$^{7}$ ~\small{Sapienza Universit\`a, Rome, Italy},
$^{8}$~\small{INFN \& Universit\`a di Pisa, Pisa, Italy},
$^{9}$~\small{ETH, Zurich, Switzerland},
$^{10}$~\small{LPNHE, Paris, France},
$^{11}$~\small{King's College London, London, UK},
$^{12}$~\small{NICBP, Tallinn, Estonia},
$^{13}$~\small{University of Amsterdam \& NIKHEF, Amsterdam, Netherlands},
$^{14}$~\small{University of Technology, Vienna, Austria},
$^{15}$~\small{LPTHE/Sorbonne Universit\'e, Paris, France},
$^{16}$~\small{Ecole Normale Sup\'erieure, Paris, France},
$^{17}$~\small{Oxford University, Oxford, UK},
$^{18}$~\small{ADAM SA, Meyrin, Switzerland}
$^{19}$~\small{Instituto de F\'isica de Cantabria, Santander, Spain},
$^{20}$~\small{Scuola Normale, Pisa, Italy},
$^{21}$~\small{Central China Normal University, Wuhan, China},
$^{22}$~\small{Boston University, Boston, USA},
$^{23}$~\small{LIP, Lisbon, Portugal}, 
$^{24}$~\small{Institute of Nuclear Physics of Polish Academy of Sciences, Krakow, Poland}
\\

\begin{center}
\bf{Abstract}
\end{center}
We review the capabilities of two projects that have been proposed as the next major European facility, for consideration in the upcoming update of the European Strategy for Particle Physics: CLIC and FCC. We focus on their physics potentials and emphasise the key differences between the linear or circular approaches. 
We stress the uniqueness of the FCC-ee programme for precision electroweak physics at the $Z$ peak and the $WW$ threshold, as well as its unequalled statistics for Higgs physics and high accuracy for observing possible new phenomena in Higgs and $Z$ decays, whereas CLIC and FCC-ee offer similar capabilities near the $t \overline t$ threshold. Whilst CLIC offers the possibility of energy upgrades to 1500 and 3000~GeV, FCC-ee paves the way for FCC-hh. The latter offers unique capabilities for making direct or indirect discoveries in a new energy range, and has the highest sensitivity to the self-couplings of the Higgs boson and any anomalous couplings. We consider the FCC programme to be the best option to maintain Europe's place at the high-energy frontier during the coming decades.

\section{Introduction}

The key task in formulating the European Strategy for Particle Physics is charting the course to the next major European facility.

Several important new facts have emerged in 2019. The CDRs of the FCC-ee and -hh projects 
were published in January~\cite{Abada2019}. In March, Japan postponed the decision about an ILC hosted 
in Japan to an indefinite date. In May, Europe discussed its particle physics strategy in 
Granada, where several high-energy options were presented. In September, the Physics Briefing 
Book~\cite{Ellis2019} was published and the European Strategy Group prepared a Supporting Note for the
Briefing Book~\cite{BriefingBookNote2019}, including five possible scenarios and raising a number 
of important issues. 

The 2013 European Strategy mandated CERN to undertake design studies for accelerator projects in a 
global context, with emphasis on proton-proton and electron-positron high-energy frontier machines. 
In parallel to these design studies, a vigorous accelerator R\&D programme should be pursued.

\section{Existing Projects}
In response to this mandate, large collaborative work programmes were organised leading to the CDRs 
for the FCC-ee and FCC-hh projects and the CLIC Project Implementation Plan (the CLIC CDR was 
published in 2011). As a result, the physics capabilities of these projects have been explored, and 
the costs of these options are also better known (see Table~\ref{tab:COSTS}).\footnote{The capabilities of the HE-LHC have also been documented~\cite{HE-LHC}, but are not discussed here.} These facilities 
require a new scale of investments, which are to be seen in the perspective of the timeframes of 
these programmes, each of which will extend over several decades, as well as the expected physics 
advances.

The Supporting Note for the Briefing Book~\cite{BriefingBookNote2019} emphasises two key issues:
\begin{itemize}
\item The adopted facility would be constructed in stages and operated over 40 or more years 
\item The facilities under consideration are either linear $e^+e^-$ colliders staged in energy, or 
circular colliders with an $e^+e^-$ collider phase followed by a hadron collider phase using the 
same infrastructure.
\end{itemize}

The Supporting Note proposes consideration of five scenarios. Besides the well-understood CLIC, ILC 
and FCC scenarios, a FCC-ERL ``green" version~\cite{Litvinenko2019} has recently been proposed. It is a linear collider
with N-pass linacs (single pass at the interaction point (IP)), energy recovery and long (about 70 km) return arcs 
between successive passes. This novel concept is complex, requires considerable further work and is 
outside the scope of this note.  We also do not consider scenario 2 of~\cite{BriefingBookNote2019},
namely building first a 
linear collider, e.g., CLIC~380, followed by the FCC-hh machine. Scenarios 4 and 5 assume that an $e^+ e^-$ machine is built in another region, which would take the lead, and are also not discussed here.

\begin{table}[!ht]
\begin{center}
\begin{tabular}{|c|c|c|c|}
\hline
 CLIC~$^{a)}$ & 380~GeV & 1500~GeV & 3000~GeV \\
 Total & 5.9 (drive beam)  & 11.0 & 18.3 \\
          & 7.3 ~~(klystron)~~~ & 12.4 & 20.1 \\
 \hline
 ILC~$^{b)}$ & 250~GeV & 500~GeV & 1000~GeV~$^{c)}$ \\
 Total & 6.0 & 8.9 & 16.4 \\
 \hline
 FCC-ee~$^{d)}$ & 250~GeV & 365~GeV & FCC-hh (100~TeV)~$^{e)}$ \\
 Total & 10.5 & 11.6 & 28.6 \\
 \hline
 CEPC & 240~GeV~$^{f)}$ & & SppC (100~TeV) \\
 Total & 5.0 & & 13.9 \\
 \hline
\end{tabular}
 \caption{\it Estimated costs (in current BCHF) of major proposed facilities.\protect\footnotemark}
\label{tab:COSTS}
\end{center}
\end{table}
\footnotetext{ These costs are total costs including the costs of previous stages but excluding experiments, taken from:\\
a)	~~~~CLIC Summary Report, CERN-2018-005-M~\cite{CLICSummaryReport}, including integrated civil engineering costs for 380, 1500, 3000 GeV (drive beam option) of 1.3, 2.3, 3.6 BCHF; \\
b)	~~~~L. Evans and S. Michizono, eds., The International Linear Collider Machine Staging Report 2017, KEK 2017-3~\cite{EM}, costs are averages of options presented, adjusted for quoted inflation of 12\%, civil engineering costs ILC 250 1.1 BCHF; \\
c)	~~~~ILC TDR Vol. 3.II~\cite{Adolphsen}, Fig. 15-16, average construction cost of 1 TeV scenarios, extrapolated from 500 GeV baseline; \\
d)	~~~~FCC-ee CDR~\cite{Abada2019}, of which the cost of infrastructure in common with FCC-hh 
  (civil engineering, electricity and ventilation) is 7.5 BCHF; \\
e)	~~~~FCC-hh CDR~\cite{Abada2019}, costs assume prior construction of FCC-ee, supplementary cost for civil engineering 0.6 BCHF; \\
 f)      ~~~~CEPC Project~\cite{CEPCCDRv1}, costs assume that local government will provide the land, and the necessary infrastructure for the CEPC facility.}

The Supporting Note~\cite{BriefingBookNote2019} has the cardinal virtue of posing directly the central
question:
\subsection{Linear or circular?}
This note addresses this question by comparing the two major projects considered for Europe at CERN,
namely CLIC~\cite{Charles2018, CLICSummaryReport} and FCC~\cite{Abada2019}. Both projects would be constructed and 
exploited in phases in order to achieve their full potential and to explore a very rich physics 
agenda, but in two quite different ways. Either facility would provide an infrastructure for 
accelerator-based particle physics in Europe for the second half of the 21st 
century~\cite{Charles2012}. Which of the two options should Europe consider? We focus here on some 
of the key scientific and technical issues selected from the wealth of available information.

\subsection{Other accelerator options}

Given that we are discussing a long-term strategy covering the next 40 - 50 years, we cannot ignore the potential of novel accelerator concepts. 

A muon collider has been discussed for many years, and has been reviewed in Ref~\cite{Delahaye2019}.
Recently, first results on muon cooling were obtained~\cite{Bogomilov2019}, and a new implementation
idea~\cite{Alesini2019} is being evaluated.  An $\mu^+\mu^-$ collider at $\sqrt{s}$ = 125 
GeV producing the Higgs boson in the $s$-channel~\cite{Rubbia2019} would make possible a scan of the Higgs boson line-shape yielding exquisite precision for its mass and 
a direct measurement of its width, but would not be competitive with FCC-ee in the measurement of 
either the width or the couplings. Conceivably, a 3 
TeV muon collider could become an attractive alternative to CLIC at 3 TeV, and it has been suggested
that a 14 TeV muon collider~\cite{Neuffer2018} might be a viable alternative to FCC-hh at 100 TeV, at
least for some physics aspects.  

\begin{figure}[ht]
\begin{center}
\includegraphics[width=8cm]{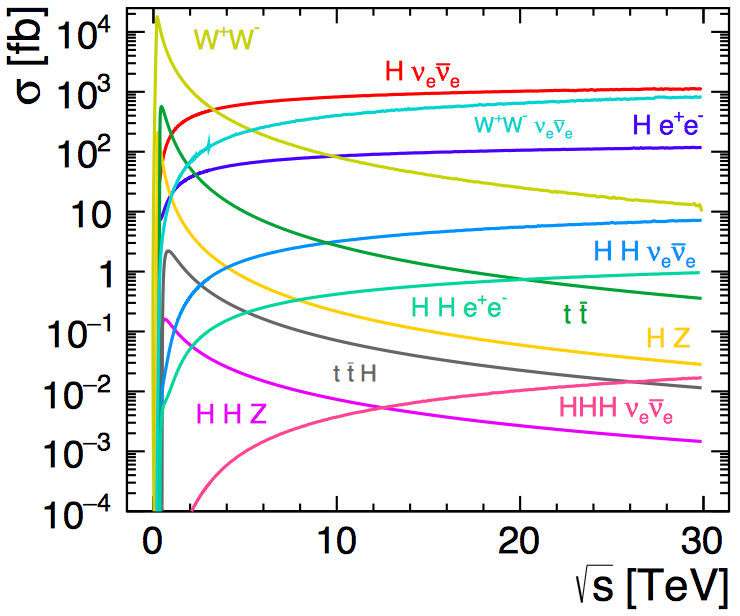}
\vspace{-2mm}
\caption{\it Production cross-sections vs. centre-of-mass energy for a lepton-antilepton collider,
evaluated at leading order and including initial-state radiation. In the case of a muon collider, $e$ and $\nu_e$ would be replaced by $\mu$ and $\nu_{\mu}$ ~\cite{CLICSummaryReport}.}
\label{fig:CrossSections1}
\end{center}
\end{figure}

However, considerable R\&D will be required to demonstrate the feasibility of a muon 
collider~\cite{Delahaye2019}, and its physics potential depends strongly on the expected luminosity.
Figure~\ref{fig:CrossSections1} shows the cross-sections of the 
main physics channels for lepton-antilepton colliders (which are compared with those for hadron 
colliders in Figure~\ref{fig:DifferentColliders}). We note that 1\% statistical accuracy on a process with 1~fb cross-section 
requires 10 years at $10^{35}$~cm$^{-2}$s$^{-1}$ luminosity, and that 0.1 fb leads to 100 events per inverse ab.

Other even more advanced schemes of acceleration, e.g., wake-field acceleration, may provide a 
breakthrough in a more distant future~\cite{Cros2019}. Both concepts require very ambitious 
accelerator R\&D, which obviously must be pursued, and the physics potential of these novel 
accelerator concepts ought to be more concretely assessed. However, as these concepts are very many 
years away from the maturity of the CLIC and FCC proposals, we do not consider a muon collider or 
new modes of acceleration in the present discussion.

In Section 3, we discuss the physics potentials of the first phases of the FCC and CLIC projects, namely electron-positron collisions at energies up to 380~GeV. In section 4, we compare the potentials at their respective ultimate stages, CLIC at 3000 GeV,  FCC-hh at 100 TeV.

\section{Stage 1: CLIC 380 or FCC-ee from the $Z$ Peak to 365 GeV}
There is overwhelming consensus on the physics agenda, with an $e^+e^-$ collider at an energy scale up to about 400~GeV, just above the $\ttbar$ threshold, as the next high-energy facility.

CLIC 380 aims at an instantaneous luminosity of 
$1.5 \times 10 ^{34}$ cm$^{-2}$s$^{-1}$ at 380~GeV, of which 60\% 
would be above 99\% of $\sqrt{s}$.
With an integrated luminosity of 1 ab$^{-1}$ about 160,000
Higgs bosons are produced at CLIC 380. This could be achieved in 3 years of luminosity ramp-up and 
5 years of data taking~\cite{Robson2018}. 

FCC-ee plans to operate at 365 GeV with a physics programme similar to CLIC 380. In addition, 
  FCC-ee offers a rich programme at the $Z$ peak, at the $WW$ threshold, and at the $HZ$ cross-section maximum.
At such energies, CLIC would have much less luminosity 
than the 1.7 ab$^{-1}$/yr @ 240 GeV and 50 ab$^{-1}$/yr of FCC-ee at the $Z$ peak (CLIC could provide
0.0025 ab$^{-1}$ per year at the $Z$ pole if unmodified, 0.045 ab$^{-1}$ if 
modified)~\cite{Charles2018Page3}. This limitation could partly be compensated, and only for 
parity-violating observables, by exploiting longitudinal beam polarisation,
provided the polarisation is known with a higher accuracy than that achieved at 
SLC~\cite{Blondel2019}. In order to make accurate measurements of the Higgs, top, electroweak (EW) and QCD sectors,
the FCC-ee programme of operations at centre-of-mass (CM) energies between the $Z$ peak and 365~GeV includes about 14 years of data taking, as seen in Table~\ref{tab:FCCeeBaseline}. 

\begin{table}[!h]
\begin{center}
\begin{tabular}{|l|c|c|c|c|c|c|}
\hline 
Working point & $Z$, {\footnotesize years 1-2} & $Z$, {\footnotesize later} & $WW$ & $HZ$ & \multicolumn{2}{|c|}{${\rm t\bar t}$} \\ 
\hline\hline
$\sqrt{s}$ {\footnotesize (GeV)} & \multicolumn{2}{|c|}{88, 91, 94} & 157, 163 & 240 & {\small 340 - 350} & 365 \\ \hline
{\small Lumi/IP {\footnotesize ($10^{34}\,{\rm cm}^{-2}{\rm s}^{-1}$)}} & 115 & 230 & 28 & 8.5 & 0.95 & 1.55 \\ \hline
{\small Lumi/year {\footnotesize (${\rm ab}^{-1}$, 2 IP)}} & 24 & 48 & 6 & 1.7 & 0.2 & 0.34 \\ \hline
Physics goal {\footnotesize (${\rm ab}^{-1}$)} & \multicolumn{2}{|c|}{150} & 10 & 5 & 0.2 & 1.5 \\ \hline
Run time {\footnotesize (year)} & 2 & 2 & 2 & 3 & 1 & 4 \\ \hline
 & \multicolumn{2}{|c|}{} & & $10^6$ \,$HZ$ & \multicolumn{2}{|c|}{$10^6 {t\bar t}$} \\
Number of events &  \multicolumn{2}{|c|}{$5\times 10^{12}$ $Z$} & $10^8$ $WW$ & + & \multicolumn{2}{|c|}{$+200$k $HZ$} \\
 & \multicolumn{2}{|c|}{} & & 25k $WW$ $\to$ $H$ & \multicolumn{2}{|c|}{$+50$k\,$WW \to H$} \\ \hline
\end{tabular} 
\caption{\it The FCC-ee baseline operation scenario~\cite{Abada2019}.}
\label{tab:FCCeeBaseline}
\end{center}
\end{table}

FCC-ee luminosity targets are 150 ab$^{-1}$ around the $Z$ peak, producing $5 \times 10^{12}$ $Z$ bosons, 
10 ab$^{-1}$ around the $WW$ threshold producing $10^8$ W pairs, 5 ab$^{-1}$ at 240 GeV, producing 
$10^6$~$HZ$ events
and 25,000 $WW \to H$ events, and 1.5 ab$^{-1}$ just above the $\ttbar$ threshold, producing $10^6~\ttbar$ 
events,  200,000 $HZ$ events, and 50,000 $WW \to H$ events. These unequalled statistics are the key to
achieving high accuracy for observing possible new phenomena, such as invisible Higgs decays, exotic $Z$, $\tau$ or top decays, lepton 
flavour violation or evidence of elusive feebly-coupled particles. 

The physics reaches of these two projects, which have been studied extensively, are summarised in the following Sections.

\subsection{Higgs couplings}
A useful benchmark for characterising the different machine sensitivities to potential deviations from the Standard Model (SM) Higgs boson properties uses multiplicative coupling strength modifiers, known as the $\kappa$ framework. The relative precisions achievable for measurements of these modifiers in the combination with HL-LHC are given in 
Table~\ref{tab:AchievableAccuracies}, together with invisible and untagged Higgs branching ratios. 
The $\kappa$ framework is suitable for spotting deviations
from the SM, but does not provide a systematic description of new physics situated at a higher energy
scale, which is better treated 
with an effective Lagrangian approach, see the very instructive reference~\cite{deBlas2019}.

\begin{table}[!ht]
\begin{center}
\begin{tabular}{|c|c|c|c|}
 \hline
 Coupling modifier & \multicolumn{2}{|c|}{HL-LHC +}  \\
 (precision in \%) & CLIC$_{380}$ & FCC-ee$_{365}$ \\
 \hline
  $\kappa_W$  & 0.73  & 0.41 \\
  $\kappa_Z$  & 0.44  & 0.17 \\
   $\kappa_g$  & 1.5  & 0.90 \\
  $\kappa_\gamma$  & 1.4~$^*$  & 1.3 \\
  $\kappa_{Z \gamma}$  & 10~$^*$  & 10~$^*$ \\
  $\kappa_c$  & 4.1  & 1.3 \\
  $\kappa_t$  & 3.2  & 3.1 \\
  $\kappa_b$  & 1.2  & 0.64 \\
  $\kappa_\mu$  & 4.4~$^*$  & 3.9 \\
   $\kappa_\tau$  & 1.4  & 0.66 \\
   \hline
BR$_{\rm inv}$ ($<$~\%, 95\% CL) & 0.63 & 0.19 \\
BR$_{\rm unt}$ ($<$~\%, 95\% CL) & 2.7 & 1.0 \\
\hline
\end{tabular}
\vspace{2mm}
\caption{\it Achievable precisions for coupling strength modifiers, adapted from Table 5 of~\cite{deBlas2019}. There is no analysis input available in the reference documentation of ref.~\cite{deBlas2019} for the entries with asterisks ($^*$), but HL-LHC results dominate the combination.}
\label{tab:AchievableAccuracies}
\end{center}
\end{table}

Whichever approach is used~\cite{deBlas2019}, FCC-ee compares favorably with CLIC 380 on the 
expected accuracy of the Higgs couplings, in particular concerning the invisible and untagged Higgs 
widths. The total Higgs width is also measured more accurately at FCC-ee than at CLIC, with precisions of 
1.0 \% and 2.7 \%, respectively, in a  global fit to the $\kappa$ parameters. 

By measuring the $HZ$ cross section at two different centre-of-mass energies (240 and 365 GeV) with high accuracy, 
FCC-ee can extract the Higgs self-coupling through its quantum effects. With 4 experiments at FCC-ee, an option under 
study, the larger data sample would yield a model-independent measurement of the Higgs self-coupling
$\kappa_{\lambda}$ with a precision of $\pm$25\%, reduced to $\pm$24\% in combination with HL-LHC.

In the FCC-ee baseline scenario, the coupling modifiers would be measured later than for CLIC, unless 
FCC-ee is run earlier at the optimal energy for Higgs physics of 240 GeV. Given the 
specificity and the relatively modest size of the radiofrequency (RF) system needed for the high-current $Z$ exposure, 
the proposed staging plan for FCC-ee, starting at the $Z$ peak, would be natural, efficient and economic, and 
hence a rational model for the machine installation and physics exploitation. 
Running first at 240 GeV is however certainly possible, but
more information is needed on such a staging scenario. 

\subsection{$Z$ and $WW$ EW programme}
Unique to FCC-ee are the very high luminosity exposures at the $Z$ peak (about $10^5$ times the LEP 
statistics of $5 \times 10^{12}$ $Z$ decays) and at the $WW$ threshold (with $10^8$ pairs of $W$ bosons to be produced), providing an 
outstanding programme of EW, QCD and
flavour physics~\cite{Blondel2016}. Such a complete programme of EW measurements is closely linked to the study of properties
of the Higgs boson, and is an essential complement to it. A high precision (<100 keV) absolute determination of the CM 
energy, thanks to 
transverse polarisation and resonant depolarisation, is an important asset and a unique feature of a circular lepton collider. A crucial input in the interpretation of the precision observables is a precise value of $\alpha_{\rm em} (M_Z)$, which can be extracted directly in dedicated runs on the edges of the $Z$ peak  from the energy dependence of the $\mu^+ \mu^-$ forward-backward asymmetry, improving the current uncertainty by a factor of 4. Forward-backward and polarisation asymmetries will allow the uncertainty in $\sin^2 \theta_W$ to be reduced by a factor of  30 to 50. The measurement of the total $Z$ width and of its visible fraction will make possible a precise extraction of the invisible $Z$ decay width corresponding to an accuracy of 0.001 in the effective number of neutrino species. The $10^8$ pairs of W bosons will enable the uncertainty in the W mass to be reduced to 0.5 MeV and that in its width to 1.2 MeV. Overall, compared to LEP, the EW observables would be measured with one to more than 
two orders 
of magnitude improvements in precision. The various measurements possible at the FCC-ee are 
summarised in Table 6 of the second reference of~\cite{Charles2012}. Longitudinal polarisation, not 
available at FCC-ee, 
would not add much that FCC-ee cannot obtain otherwise, since helicity effects in the final state, 
e.g., of the $\tau$ 
in $Z\to \tau^+\tau^-$, will provide similar information. A dedicated theoretical effort is envisaged~\cite{Blondel2016}, to match 
the expected experimental accuracy and to make FCC-ee the ultimate EW and flavour factory.
\subsection{Top physics}

The luminosities of CLIC near the nominal CM energy of 380~GeV and that of FCC-ee   per IP at 365 GeV are comparable, so the 
performances of the two machines are \emph{a priori} similar, and their measurements near threshold will bring much information on the top quark mass and couplings. 
Both machines provide a very pure sample of $\ttbar$ events to measure the cross-section at each 
energy. The top mass is extracted from the energy dependence of the cross-section. Statistical 
uncertainties (17 MeV at FCC-ee, 20 MeV at CLIC-380) are similar. The total uncertainty is about 50 
MeV, presently dominated by the scale uncertainties of the NNNLO QCD prediction for the top 
threshold region~\cite{Beneke2015}.
The sensitivity of the FCC-ee~\cite{Janot2015} to the top quark electroweak couplings is excellent 
for a CM energy just above the $\ttbar$ threshold, with an optimum at 365 GeV. The lack of beam 
polarisation is compensated by measuring the polarisation of the top quarks, which is transferred to
their decay products via the weak decay $t\to Wb$ and is measured in their distributions, e.g., those
of the leptons from W decays. With conservative assumptions on lepton identification, b-tagging 
efficiencies, and lepton angular and momentum resolutions, 
after three FCC-ee years at 365 GeV, one will reach absolute precisions of the order of $10^{-3}$ ($10^{-2}$) for $A_\gamma (B_\gamma)$ and $3\times 10^{-3}$ ($2 \times 10^{-2}$) for $A_Z (B_Z)$, where $A_V$ and $B_V$ are the vector and axial SM top couplings to a boson $V$
with $V=\gamma, Z$ (see~\cite{Contino2016} for details).
The experimental accuracy will be exploitable, if the theoretical prediction for 
the top-pair cross-section reaches an accuracy below 2\%, a challenge only 20 GeV above threshold, but 
achievable. With the expected accuracy on these couplings, the $\ttbar$H coupling can be extracted
at the FCC-hh with sub-per-cent-level accuracy~\cite{Contino2016}.

The FCC-ee has access to the top Yukawa coupling on its own, through its effect at the quantum level
on the $\ttbar$ cross-section just above production threshold. The FCC-ee measurements at lower 
energies are important to fix the value of the strong coupling constant $\alpha_s$. This precise 
measurement allows the QCD effects to be disentangled from those of the top Yukawa coupling at the $\ttbar$ vertex. 
A precision of 10\% on the top Yukawa coupling is thus achievable at the FCC-ee (see the FCC-ee Section in~\cite{Abada2019}).
This coupling will, however, be determined with a similar or better precision 
already by the HL-LHC (3.4\%, model dependent~\cite{Cepeda2018}), and constrained to 3.3\% through a 
combined model-independent fit with FCC-ee data.

Neither FCC-ee nor CLIC 380 can access $\ttbar X$ states, a handicap compared to the CLIC 1.5 / 3 TeV 
upgrades. CLIC has released a global interpretation of top-quark pair production using an effective field theory framework involving seven independent operators, and requiring at least two energy stages~\cite{Abramowicz2018}. The first phase alone 
gives access to new physics scales of more than 10 TeV. But the real breakthrough in sensitivity 
would require the higher energies, as shown in~\cite{Abramowicz2018}.

We conclude that for top physics the FCC-ee and CLIC 380 programmes are approximately equivalent.
However, FCC-ee offers a wider research programme and powerful physics reach due to the high 
luminosity at CM energies below 365 GeV and 2-experiment operation. This is a significant advantage
compared to CLIC 380.

\subsection{General Issues for CLIC 380 and FCC-ee 365}
CLIC 380 requires a linear tunnel of 11.4 km length. Presently, two options for acceleration - 
drive-beam acceleration or X-band RF klystrons - are being considered, with a power consumption of 
170 MW in both cases.  The vertical beam size at final focus is 2.9 nm~\cite{CLICSummaryReport}. 
The cost for CLIC 380 
ranges from 5.9 BCHF (drive beam option) to 7.3 BCHF (klystron option), with an uncertainty of about 1.5 
BCHF.

Achieving the luminosity, and hence the quality of the measurements, needs control of beam size, 
machine parameters and stability at the nm scale, e.g., mechanical stabilisation of the final focus 
quadrupoles to 0.3 nm r.m.s.. This is the major technological challenge and concern. Imperfections 
(misalignments, ground motion, vibrations) in the accelerator complex are the main limit to the 
luminosity. The very ambitious CLIC parameters are backed up by simulation studies, measured 
hardware performances and beam tests, indicating that the design parameters could be achieved. 
However, the critical issues have only been tested individually, but not yet together in an
integrated 
set up. A risk remains, given the extreme level of performance required from the full complex. The 
only linear $e^+ e^-$ collider ever built, the SLC, achieved 40\% of its design luminosity in its best, 
tenth and final year, with a vertical size at the IP of about 600 nm and luminosity 
$3 \times 10^{30}$ cm$^{-2} $s$^{-1}$. These performance limitations may be linked to the sub-optimal 
topological features of the final arcs. 
The reliability of CLIC results must be achieved with a single, albeit fully optimised, detector.
Obviously, one can devise innovative procedures to ensure cross-checks of the results, e.g., two 
collaborations exploiting independently the same detector, but with common hardware.
Still, aiming at 10 to 100 times better accuracy will open a
totally new range of systematic errors and pitfalls, never encountered before.

FCC-ee requires a circular tunnel of approximately 100 km circumference, a very big enterprise, 
but one that would provide an invaluable infrastructure, offering the prospective of a long-range 
and diverse physics programme, extending over many decades: this is the perspective of the 
integrated programme of FCC-ee followed by FCC-hh~\cite{Charles2012}. The cost of the civil
engineering for the FCC-ee programme is presently estimated to be 5.4 BCHF. Optimisation of the
circumference, within $\pm10$
\%, needs to be done to minimise the cost per km. The cost for civil engineering
covering both FCC-ee and FCC-hh with the 4 caverns presently considered (2 large and  2 smaller caverns are foreseen) will require an additional 0.6 BCHF. The 
complete FCC-ee programme with two experimental caverns will require a 
total investment estimated at 11.6 BCHF. The cost of a possible initial phase at 250 GeV is
estimated at 10.5 BCHF.

The tunnel represents an up-front expenditure and needs to be considered in the timeframe of the 
integrated programme. Given the duration of the construction over many years, the expenditure will 
be staged. Nevertheless, major changes will be needed with respect to the present CERN way of 
programme realisation: exceptional and quite substantial contributions from the host states, 
significant contributions from non-member states participating in this programme and preferential
loans with long-term reimbursement profiles. We assume that this should be possible.

This circular machine is also quite demanding, given the novel features involved (top-up injection, 
low emittance requirement, high beam current at low energy, "tapering" at high energy, etc.). For 
the $Z$ run, the crab waist scheme contributes to boosting the luminosity and reaching over
four orders of 
magnitude more than LEP. However, the machine will be built using the vast 
experience accumulated with previous circular $e^+e^-$ colliders, from LEP200 at high energies
$\le 209.2$~GeV and PEP-II, with its high beam current, to the recent 
advances at the Super KEK B-factory and at Da$\Phi$ne in Frascati. Past circular $e^+ e^-$ colliders have 
typically achieved their design luminosities within a few years and subsequently exceeded them (e.g., by a 
factor 4 in luminosity at LEP2).

{In summary}: FCC-ee, with two experiments simultaneously taking data, offers the more 
attractive physics reach, but is more expensive than CLIC 380. The cost is higher, because part of 
the "upgrade" - e.g., some infrastructure and civil engineering for FCC-hh - is included. The attractiveness becomes even 
more compelling - both physics- and cost-wise - when considering the integrated combination with 
FCC-hh compared to CLIC 3000.

\section{Stage 2: CLIC 1500/ 3000 or FCC-hh}

The possibility to increase the CM energy in stages is a key advantage of a linear $e^+e^-$ 
collider. In the case of the CLIC 1500 (5.1 BCHF additional cost, 29 km tunnel, 364 MW) 
and CLIC 3000 (7.3 BCHF added, 50 km 
tunnel, 18.1 BCHF total, 589 MW) options, a discovery is always possible, although
there is 
presently no indication of new physics (supersymmetry, compositeness, dark matter or other new particles) in
this
energy range. The HL-LHC and possibly other programmes will tell us more and, if nothing appears, 
indicate whether any notorious "blind spots" remain unexplored.

The physics case for CLIC 1500 and 3000 is thus mostly motivated by their programmes of measurements
in the Higgs, top, QCD and EW sectors. These stages are crucial for precision measurements and the 
list of potential accuracies is considerable.

An assessment of the actual accuracy of these measurements and of their reliability is again the key
issue. The expected quality of results implies reaching the design luminosity in a reasonable time span, 
requiring a vertical size of the collision region at the nanometer level, hence small vertical 
emittance and strong vertical focusing. As for CLIC-380, the reliability of the physics results must
be guaranteed despite using a single detector.

The 100 TeV FCC-hh, presently estimated at an additional 17 BCHF with infrastructure inherited from
FCC-ee, will 
represent a major step in energy compared to LHC. As for CLIC 3000, the FCC-hh power 
requirement, presently evaluated as being close to 600 MW, obviously needs a special, environmentally sound and probably 
radical solution, which must be considered in the overall project planning. Uniquely, the FCC-hh programme 
includes ion-ion and possibly electron-hadron collisions, which offer new insights into the collective behaviour of hadronic matter such as the quark-gluon plasma and the colour-glass condensate, but here we concentrate on the pp programme.

FCC-hh offers the possibility of discoveries in an unchartered mass range~\cite{Contino2016} and with its  planned integrated
luminosity of 30~ab$^{-1}$
guarantees huge yields of known objects, e.g., $\sim 3 \times 10^{10}$ Higgs produced, $\sim 4 \times 10^7$ Higgs 
pairs, $10^{13}$ top quark pairs,
etc., a totally new class of measurements. In many cases, such as the Higgs self-coupling, it
offers the best available capability. The increase in production cross-section with respect to the
LHC ranges from a factor of $\approx$10 for $VH (V = W, Z)$ associated production to a factor of
$\approx$ 60 for the 
$\ttbar$H channel, and FCC-hh will provide six times more instantaneous luminosity even than HL-LHC. 
Such large statistics allow for precision in new kinematic regimes and give access to rare decay 
channels, complementary to FCC-ee (see below). 
Because of the larger phase space, the rate increase is much higher for large transverse momentum phenomena, which are particularly interesting for probing heavy new physics.
The most remarkable feature of Higgs production at 100 TeV is not just the rate increase w.r.t. LHC,
but the extreme kinematical range over which production of the Higgs boson and the top quark can be explored.
The inclusive Higgs cross-section will thus be accessible experimentally at percent level accuracy, 
making possible an in-depth theoretical understanding of this process, necessary for a successful programme
of Higgs phenomenology at FCC.

\subsection{Top physics}
The top Yukawa coupling is one of two key parameters required for the understanding of the Higgs 
potential, and is crucial for the measurement of the Higgs self-coupling. The top Yukawa coupling 
determination at HL-LHC will be limited to a model-dependent accuracy of around 3.4\% because of 
statistical and theoretical uncertainties.

An effective field theory (EFT) analysis of semi-leptonic $\ttbar$ production at all CLIC stages~\cite{Abramowicz2018} in a 
global fit gives the expected sensitivities for Standard Model EFT operator coefficients. 
Operation at high 
energy improves the sensitivities to the 4-fermion operator coefficients, which approach the level 
of $10^{-4}$ TeV$^{-2}$.
At CLIC 1500, $\ttbar$H production gives direct access to the top-quark Yukawa coupling with an 
expected accuracy of 2.9\%. The $\ttbar$H process is also sensitive to a CP-odd contribution to the 
$\ttbar$H coupling~\cite{deBlas2019}.

At FCC-hh, new possibilities will be offered by the combination with the measurements of top 
properties and of Higgs branching ratios at FCC-ee, the large $\ttbar$H and $\ttbar$Z production 
cross-sections at the pp collider, and the cancellation of the dominant theory uncertainties in 
their ratios will offer new prospects. As an example, the top-quark Yukawa coupling will be inferred
at FCC with an accuracy of about 1.5\%~\cite{Mangano2016}.

\subsection{Higgs physics}
\subsubsection{Higgs couplings}
\begin{figure}[ht]
\begin{center}
\includegraphics[width=8cm]{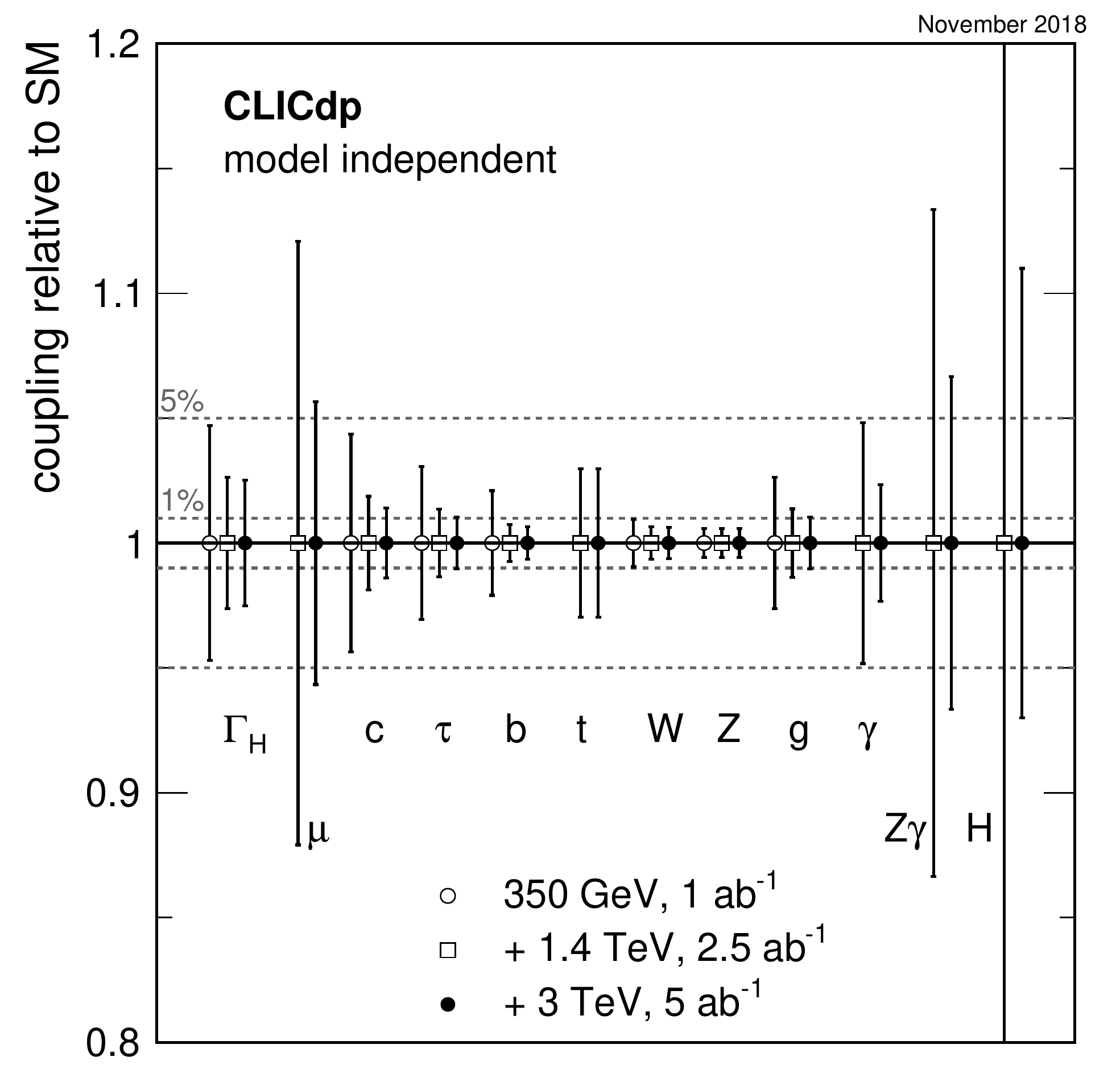}
\vspace{9mm}
\includegraphics[width=14cm]{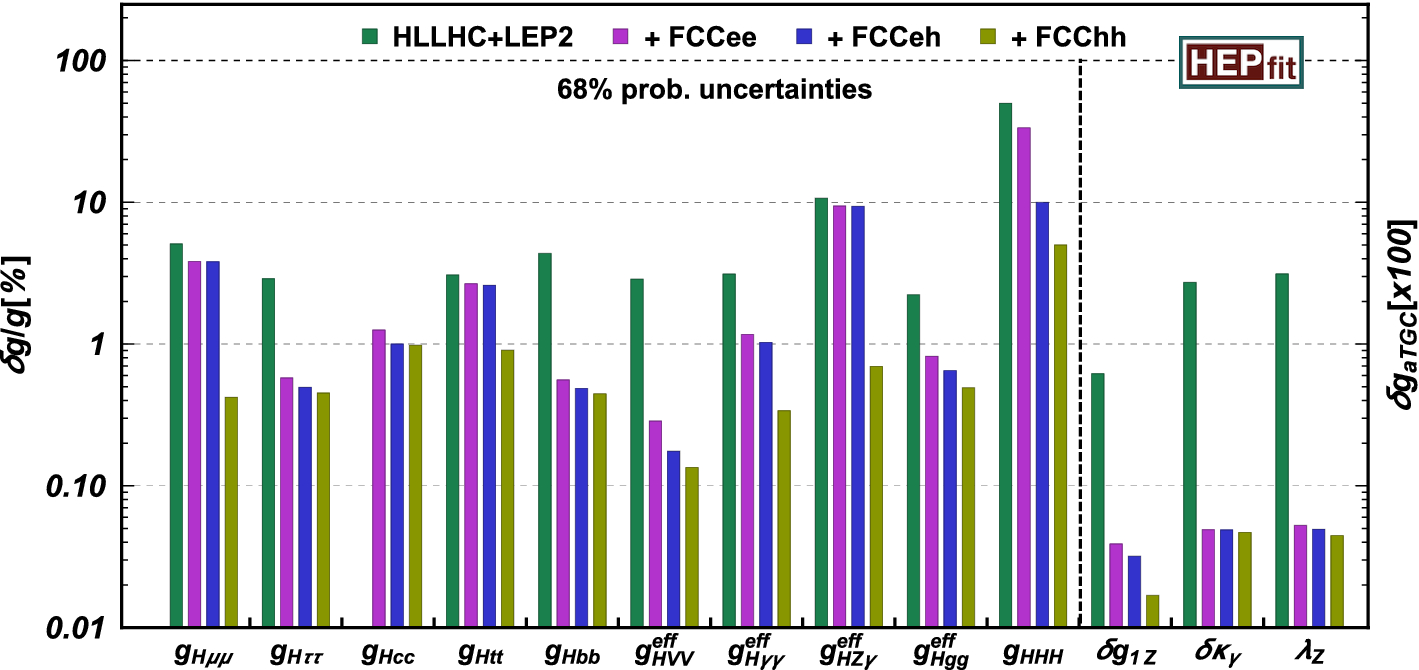}
\vspace{-6mm}
\caption{\it Accuracies in measurements of Higgs couplings relative to SM values for the 3 energy stages of CLIC~\cite{Robson2018} (upper), and for the various stages of the FCC programme~\cite{Abada2019} (lower) including HL-LHC and measurements from LEP2.}
\label{fig:Accuracies}
\end{center}
\end{figure}

At CLIC, a model-independent (MI) global fit at the three energy stages gives the results of 
Fig.~\ref{fig:Accuracies} (top)~\cite{Robson2018}, which shows the gain brought by the high-energy 
options. One expects an accuracy on $g_{HZZ}$ of 0.6\% from the total $HZ$ cross-section. 
The precision for other 
couplings such as $g_{HWW}$ and $g_{Hbb}$ reach a similar level. The $g_{Hcc}$ coupling, which is 
challenging at 
hadron colliders, can be obtained with percent-level accuracy and the total Higgs width with 2.5\% 
accuracy. Assuming the absence of non-SM Higgs boson decays, a global fit, model-dependent and 
equivalent to the hadron collider approach, constrains several Higgs couplings  to per mille-level 
accuracy thanks to the high-energy stages.

At FCC-hh, the large statistics allow for precision in new kinematic regimes and give access to rare
decay channels, complementary to FCC-ee. Percent or sub-percent accuracy on "rare" couplings, such
as $H \to \mu^+\mu^-$, $\gamma\gamma$ and $Z\gamma$,
is within reach when combined with the absolute $HZZ$ measurement in $e^+e^-$, see Fig.~\ref{fig:Accuracies} (lower).

\subsubsection{Trilinear Higgs self-coupling (SC)}
The Higgs SC is a most crucial measurement, and deviations from the SM expectation can reach tens of
percent in some new physics scenarios. 

At the LHC, ATLAS and CMS in combination expect to find evidence for $HH$ production at the 4-$\sigma$ level, and
to determine $\lambda$ with an uncertainty of 50\%, if it takes its SM value.
This may be improved further by dedicated analyses aimed at constraining $\lambda$~\cite{Cepeda2018}.

Both CLIC at high energy and FCC-hh give access to the SC at tree level through double Higgs 
production, and FCC-hh also opens the way towards a measurement of the Higgs quartic coupling.

CLIC 1500~\cite{Roloff2019} makes possible a 5$\sigma$ observation of the double Higgsstrahlung process 
$e^+e^-\to ZHH$ and provides evidence for the $WW$ fusion process $e^+e^-\to HH\nu_e\nu_e$  at the 
3.6$\sigma$ level, if $\lambda$ has its SM value. These measurements are complementary since, with 
$\lambda$ at the level of the SM value, the $ZHH$ cross-section increases, while $HH\nu_e\nu_e$ 
decreases. According to~\cite{Roloff2019}, CLIC 3000 with 5 ab$^{-1}$ would provide $\sim 370$ reconstructed $HH\nu_e\nu_e$ events with an estimated signal-to-background ratio of about 1:1. At this energy, $WW$ fusion is the leading double-Higgs production mechanism, and makes 
possible measurements of differential distributions for $HH\nu_e\nu_e$ that improve further the knowledge of
the Higgs self-coupling. Overall one expects a precision on the Higgs self-coupling $\lambda$ of 
[-7\%, +11\%] if it has the SM value.
These accuracies in the Higgs SC measurement and the possible sizes of deviations in some extensions
of the SM are strong motivations to operate CLIC at high energy. 
The equivalent event numbers expected at the ILC
are modest,\footnote{It is estimated~\cite{Tian2013} that the ILC would yield $\sim 35$ reconstructed events at 1000~GeV and $\sim 8$ reconstructed events at 500~GeV, in both cases with an integrated luminosity of 2 ab$^{-1}$ and an estimated signal-to-background ratio of about 1:1.}, so it would be crucial to 
reach the design luminosity.

At FCC-hh, the combination of the high rate for the gluon-fusion mode of double-Higgs production (a factor 40 more than
at LHC) and high luminosity (another order of magnitude more than HL-LHC) lends itself to the exploitation of several final states. Using the main 
$HH\to \bbbar \gamma\gamma$ channel plus a few secondary ones, and with the baseline detector 
performances, a precision on the Higgs self-coupling of about 5\% appears 
achievable~\cite{Yao2013}. For the main channel and 30 ab$^{-1}$, two different analyses offer 12 k 
(4 k) signal events for 27.1 k (7.3 k) backgrounds (statistical error only, see~\cite{Yao2013} for a 
discussion of systematics). Such a measurement would explore classes of models relying on modifying 
the Higgs potential to drive a strong first-order EWSB phase transition, which is necessary for scenarios of electroweak 
baryogenesis.

FCC-hh is also the only machine giving access to the quartic Higgs coupling, either via loop effects on di-Higgs production or directly via triple Higgs production~\cite{Yao2013}.

\subsection{Searches for new physics (NP)}
Indirect indications of possible NP can be provided by a variety of accurate measurements of known processes, while a direct indication of NP requires the proof of existence of a new particle or mechanism.

Both machines will address several of the major, fundamental open questions of particle physics, as 
documented extensively in~\cite{Contino2016, deBlas2018, Golling2016}. Examples are the 
possible composite nature of the Higgs, solutions to the hierarchy problem, baryogenesis and the 
electroweak phase transition, the nature of dark matter, the origin of neutrino mass, and the structure
of flavour-changing neutral currents (FCNCs).

FCC-hh will have extensive capabilities for all of these questions. CLIC will certainly have 
interesting capabilities for detecting new particles with TeV-scale masses, though many of the 
possibilities will have been explored previously by HL-LHC.

\subsubsection{Indirect discoveries}
\begin{figure}[ht]
\begin{center}
\vspace{-1.0cm}
\includegraphics[width=14cm]{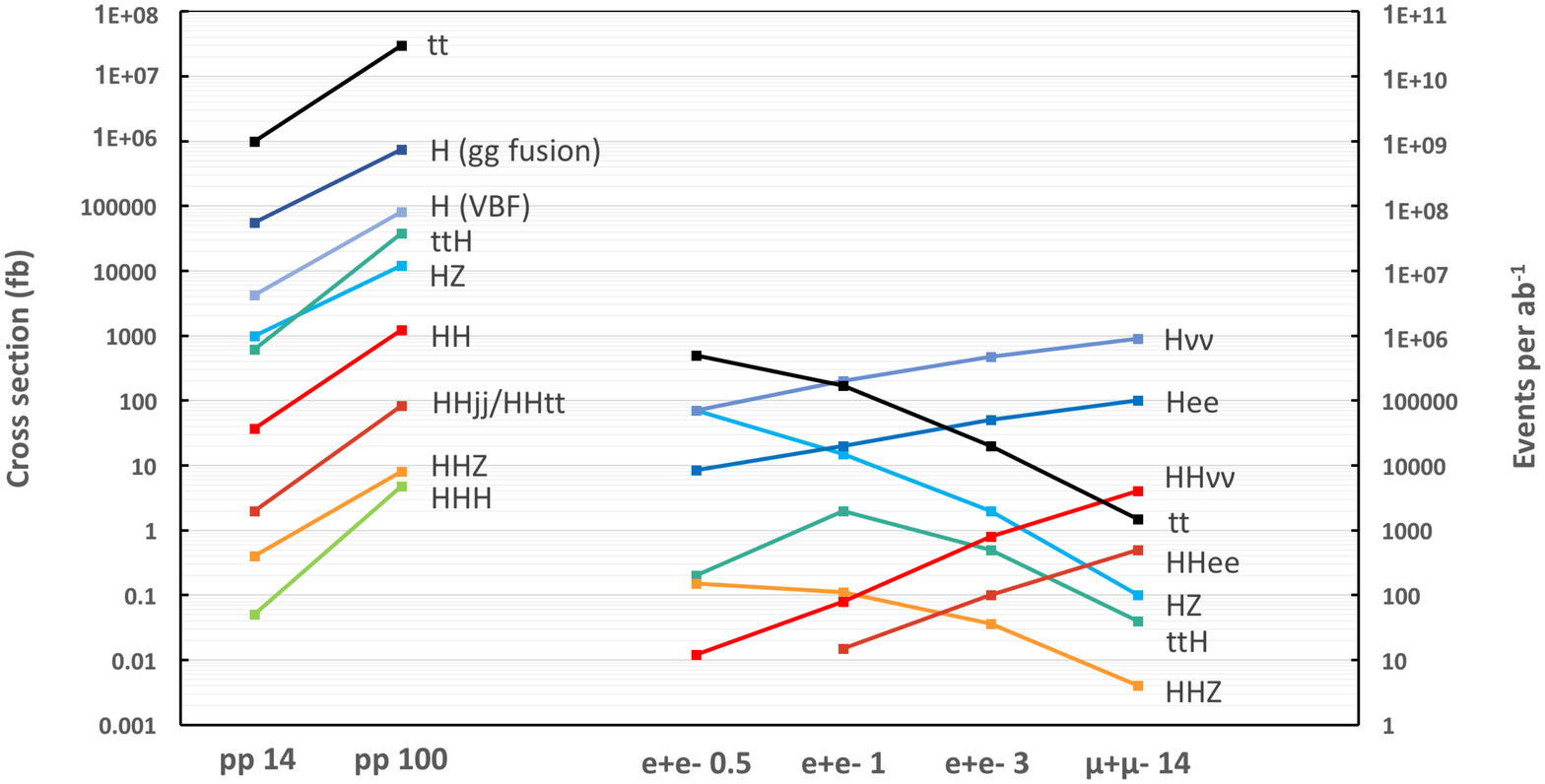}
\vspace{-7mm}
\caption{\it Some representative cross-sections at different colliders, with energies in TeV.
The $pp$ cross-sections are from~\cite{Contino2016, Baglio2016}, the $e^+ e^-$ cross sections are from~\cite{CLICSummaryReport, Charles2018, Robson2018}, and the $\mu^+ \mu^-$ cross sections are from~\cite{Delahaye2019}.
According to~\cite{Tian2013}, double beam polarization can increase the $HHZ$ and $HH \nu \nu$ cross-sections by a factor of 1.5 - 2.}
\label{fig:DifferentColliders}
\end{center}
\end{figure}

The sensitivity to indirect discoveries is determined by the production rate of known processes. 
Some representative cross-sections at the two machines, in particular for various multi-boson final 
states, are given in Fig.~\ref{fig:DifferentColliders}, which has been drawn 
from~\cite{CLICSummaryReport, Charles2018, Robson2018, Delahaye2019, Contino2016, Tian2013, Baglio2016}. In order to fix ideas about
rates, one may recall that a cross-section of 0.1 fb leads to 100 events per inverse ab, which must be 
multiplied by the appropriate branching ratios, efficiencies and cuts. For the leptonic machines and
a specific process X, one can distinguish annihilation processes (with cross-sections decreasing as 
1/s) from W-boson fusion processes (whose cross-sections increase logarithmically with s), which 
gives $\nunubar$ + X final states. The smallness of most cross-section limits the statistics 
available with CLIC, whereas FCC-hh must extract the relevant signal from high backgrounds.

High-energy vector boson scattering, where the Higgs boson plays a key role in controlling the rate 
increase, will be studied at both machines. At FCC-hh, cross-sections with longitudinal 
polarisations $V_LV_L$ can be measured with 3-4\% accuracy (e.g., using $W_LW_L$ same-sign dileptons), 
providing a measurement of the $HWW$ coupling at the percent level and a direct unitarity test of the 
amplitude for $W_LW_L$ scattering at very high energies. As another example, CLIC can discover 
indirectly a $Z'$ with SM couplings and a mass up to 20 TeV, whereas FCC-hh can discover such a $Z'$ 
directly. FCC-hh can also explore compelling classes of models with a 1st-order EW phase transition,
both directly, by looking for states required to modify the transition (which occurred at 
temperatures near the weak scale), and indirectly, e.g., by measuring the Higgs SC, as previously 
mentioned. With the accuracies expected for all the measurements and with no deviations observed 
with respect to the SM, a global fit to all FCC results will set constraints on NP up
to a scale of $\sim$ 20 to 100 TeV, assuming unit-strength tree-level
couplings, and on NP coupled to the Higgs sector up to $\sim$ 10 TeV. This sensitivity to 
new physics matches and even extends the direct discovery potential of the FCC-hh.  For comparison, CLIC can probe 
the Higgs (top) compositeness scale up to 10 (8) TeV.

\subsubsection{Direct discoveries}
CLIC exploits the cleanliness of events, within its mass and rate limits. FCC-hh benefits from the 
huge production rates of potential new objects, when gluons dominate their production, and of 
its kinematical reach, but must beat the backgrounds. We give only a few examples.

CLIC 3000 can probe the existence of new particles interacting with the SM with EW-sized couplings, up to
its kinematic limit of 1.5 (3.0) TeV, if they are pairwise (singly) produced. A prototypical example
of an extended Higgs sector is the extension of the SM with a new scalar. Particularly challenging 
is the case of a scalar without gauge interactions and interacting with the SM only through the
Higgs 
boson portal. In the case of supersymmetry (SUSY), CLIC can discover a thermal Higgsino DM candidate up to 1.2 TeV, 
a slepton up to 1.5 TeV, and the NMSSM scalar singlet mass can be probed up to 1.5 TeV~\cite{Charles2018}. Regarding 
the possibility of EW baryogenesis~\cite{RT}, the required existence of new scalar particles coupled with the 
Higgs can be exhaustively probed at CLIC by direct searches, complementing its precise measurements 
of Higgs couplings.

The 100 TeV FCC-hh offers important and unique physics opportunities, especially in connection to 
Higgs physics and its discovery potential for BSM Higgs sectors. One of the main reasons for 
increasing the CM energy is obviously to extend the reach for direct production of new heavy 
states up to tens of TeV, e.g., a SM-like $Z^\prime$ (SSM),
a $Z^\prime$ in a left-right model (LRM) or a scenario with extra dimensions, or a new charged vector boson W$^\prime$,
with a mass up to 40 TeV. However, equally 
important is the huge increase in the production rate of light states like the Higgs, which allows 
detecting exotic Higgs decays with tiny branching ratios smaller than $10^{-8}$, provided that the 
final state can be extracted from the background. In general, a 100 TeV proton-proton collider would
be sensitive to EW-charged BSM states with masses of 5-10 TeV. It would enable probes of the most 
challenging aspect of the TeV scale, namely SM singlets. It would offer high-precision high-energy 
measurements, e.g., of Drell-Yan dilepton spectra, which are sensitive to new EW-charged
states with masses around a TeV.
In the case of top quark FCNC searches  the FCC-hh can offer two orders of magnitude 
more sensitivity than the HL-LHC.

\begin{figure}[ht]
\begin{center}
\begin{minipage}{0.53\linewidth}
\includegraphics[height=5.7cm]{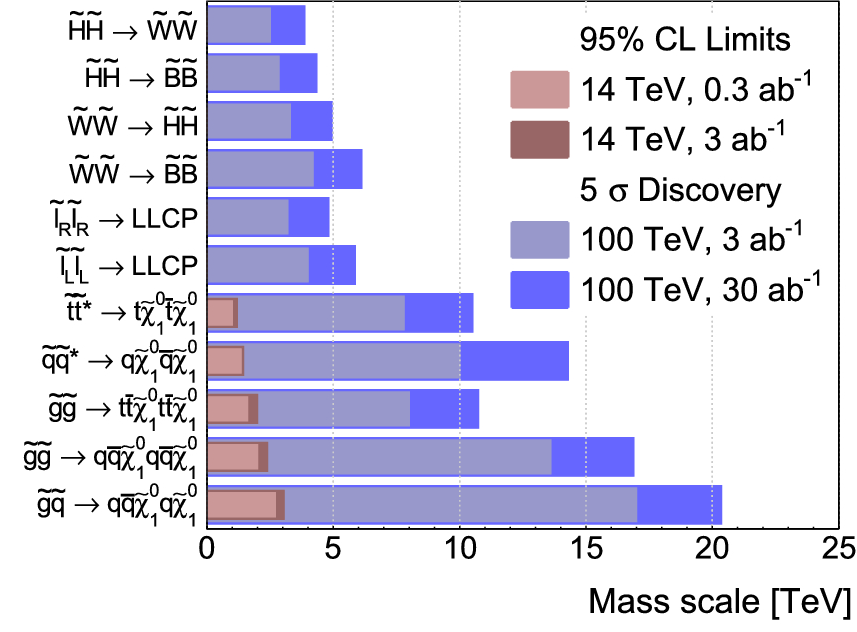}
\end{minipage}
\hfill
\begin{minipage}{0.43\linewidth}
\includegraphics[height=5.7cm]{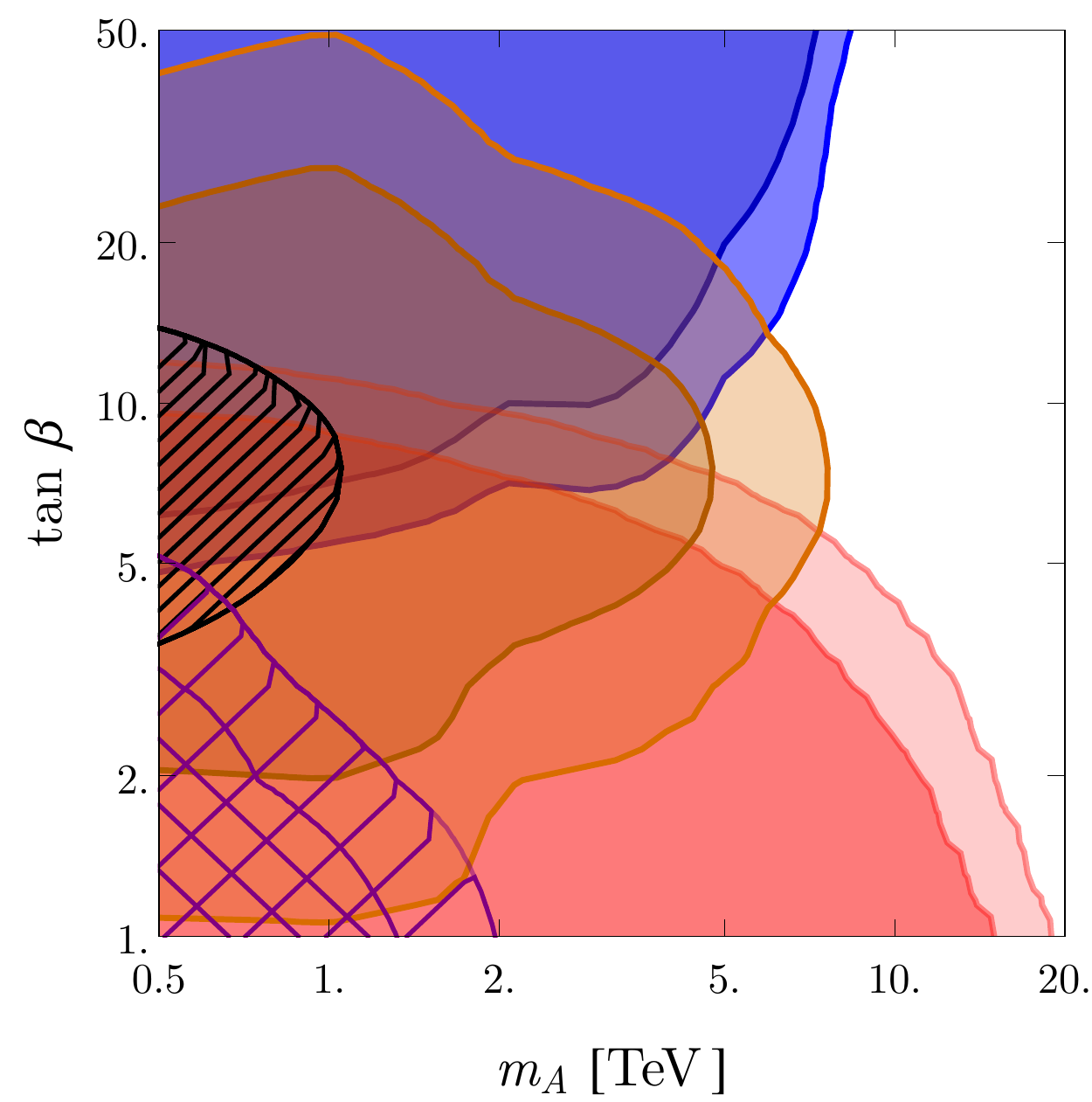}
\end{minipage}
\caption{\it FCC-hh discovery ranges for sparticles~\cite{Abada2019} (left panel) and for heavy MSSM Higgs bosons~\cite{Craig2017} (right panel)}
\label{fig:ExclusionLimits}
\end{center}
\end{figure}

SUSY, although still elusive, has at least passed one of its crucial tests: the existence of a light 
SM-like Higgs boson. As seen in Fig.~\ref{fig:ExclusionLimits},
the mass reach for gluinos at FCC-hh varies from 11 TeV (if gluinos decay mainly to 
top+antitop+neutralino) to 21 TeV (if the gluinos decay mainly to quark-antiquark+neutralino). The value of the Higgs
mass suggests that, except for small $\tan\beta$, a stop squark may be within the FCC-hh 
range~\cite{Badziak2015}. We recall also that SUSY and many other models feature larger Higgs sectors 
than the single Higgs in the SM. Direct production of additional TeV-scale Higgs states will be a
major physics goal of a 100 TeV collider with, in the MSSM case, the mass reach shown in 
Fig.~\ref{fig:ExclusionLimits}~\cite{Craig2017}. It illustrates the high sensitivity also for small 
$\tan\beta$, which would constrain the model dramatically if there were no discovery.

As for DM and other invisible states, the 100 TeV collider will be sensitive to scalar mediators as 
well as DM masses of more than a TeV. With 3000 fb$^{-1}$ of data it is expected to discover or 
exclude Higgsino thermal dark matter candidates with masses $\sim$ 1 TeV, if Wino NLSPs are lighter than 
about 3.2 TeV.  With 30 ab$^{-1}$, the reach is 1.4 TeV for pure Higgsino DM and 6 TeV for pure Wino
DM. 

Concerning the origin of $\nu$ mass, variations on the usual see-saw mechanism, such as low-scale 
see-saw scenarios or radiative $\nu$ mass models, stipulate new fields at the electroweak or 
TeV scale, whose 
direct production may be accessible at the FCC-ee or FCC-hh respectively.

\subsection{Overview of the FCC programme}
In the search for new physics at high masses, the physics potential of CLIC 1500 or 3000 would be significantly better 
than that of a 365 GeV FCC-ee. Therefore, CLIC's main interest rests upon increases in its CM energy, 
which offer a very interesting physics programme, but CLIC will forever be restricted to $e^+ e^-$ collisions. 
Given the cost and effort needed for CLIC, it will very likely preclude Europe from 
pursuing hadron collider physics beyond the LHC. This fact, added to our previous remarks about its physics
performance, the risks of the machine and the limitation to a single detector are the issues to be 
assessed before CLIC could become a viable recommendation as the prime project for the European Strategy.

In our view, the integrated FCC programme, with its combination of both $e^+e^-$ and hh collision modes,
has a much larger physics potential than 
CLIC. This programme will offer both ``guaranteed deliverables", i.e., high-quality measurements in 
all sectors 
at both the ee and pp stages, and an unrivalled discovery potential, with an increased reach for 
direct discovery at the highest masses, accompanied by an increased sensitivity in the whole mass 
range, because of higher precision and statistics. The combination of FCC-ee and FC-hh will
provide a forefront scientific programme for CERN for many decades, just as the combination of
LEP of LHC has done. We consider FCC to be a visionary programme for the future of CERN.

Concerning the top quark, the combination of the measurements of its properties and of Higgs 
branching ratios at FCC-ee with abundant $\mttbar H$ and $\mttbar Z$ production at FCC-hh will allow
the top Yukawa coupling to be measured at percent-level accuracy.

Concerning the Higgs boson, the FCC-ee results (e.g., for the total $H$width and the absolute 
$g_{HZZ}$ determination) will provide the inputs necessary to exploit fully the potential accuracy
of systematics-dominated FCC-hh measurements, such as the branching ratios for rare decay modes of 
the Higgs boson or its self-coupling. 

With the accuracies estimated for all measurements and if no deviations from the SM are observed, a 
global fit will set constraints on new physics up to a scale of $\sim$ 100 TeV
(for unit coupling), and on new
physics coupled to the Higgs sector up to $\sim$ 10 TeV. 

FCC-hh will push to the highest masses the direct search for new particles, with a reach that 
promises to confirm or exclude in a conclusive way many scenarios, such as the existence of WIMP 
DM candidates or the strong first-order nature of the EW phase transition.

If relatively light objects exist at the TeV scale with sufficiently weak interactions, they 
should be discovered at the 100 TeV collider. This includes some singlet scalar states, or an EW 
multiplet whose neutral component contributes substantially to the DM relic density. In that sense, 
the FCC-hh acts as an \emph{intensity frontier experiment} for uncoloured physics near the TeV 
scale~\cite{Arkani2016}.

Another advantage of hadron collisions, which was not a priori obvious, but has been demonstrated by
the quality of the LHC detectors, is the ability to exploit the abundance of data in a wide variety 
of channels, including in particular the spectroscopy of low-mass states.

FCC-hh should maintain 100 TeV as its ultimate target. Lower energies would weaken the physics case 
and proceeding via a lower-energy option would add significantly to the total cost. However, achieving the ambitious 100~TeV goal will require pursuing a sustained  programme to develop high-field collider magnets.

These then are some of the perspectives for the integrated FCC-ee and FCC-hh programme~\cite{Abada2019}:
\begin{enumerate}
\item \emph{Origin of the Higgs field}: this necessarily implies a new level of physics beyond the 
SM. The FCC programme is the only way to unravel its fundamental origin.
\item \emph{Dark Matter}: Although extensively studied, little is known. The neutral, spin-0 
Higgs may be the only particle coupling to a hidden particle sector, which does not interact with SM
particles. A large class of such models, providing DM candidates, would be explored at the FCC.  
\item \emph{Baryogenesis}: This is linked to the cosmological matter-antimatter asymmetry, which the SM 
alone cannot explain, and may be related to the EW phase transition (PT). A necessary
but not sufficient condition for electroweak baryogenesis 
is a first-order PT, which should have occurred at some temperature near the EW scale. New 
states responsible for driving this PT are expected to have masses near the EW scale. Numerous 
available models could be tested at the FCC, which could probe the first order nature of the
PT.
\item \emph{Neutrino masses}: Again, the SM alone cannot account for them. A plethora of
explanations could
be explored at the FCC, via both precision measurements and searches for direct production.  
\end{enumerate}

\section{Keeping HEP at the high-energy frontier vibrant}
In many research fields -- particle physics, astrophysics, planetary and space research -- projects 
may take 20 or more years from conception to archiving legacy data.  This situation requires novel 
ways for guiding the careers of young researchers, students, junior and senior post-doctoral researchers~\cite{BGINS}. Providing 
a broad education, stimulating and challenging research topics, creative ways for assuring 
appropriate visibility and recognition in multi-thousand-member research teams, and developing 
competences that are highly valued outside academia, are just some of the issues to be addressed in 
a fresh and determined way. CERN's unique infrastructure, in combination with a diverse research programme 
away from the high-energy frontier,  will allow young researchers to participate 
in more than one programme, e.g., working on an FCC research topic in one group, while in parallel participating in the construction or analysis of an ongoing experiment, or carrying out detector R\&D. The senior community has the responsibility to provide the 
best environment for the upcoming generation. Personal research ambitions need to be balanced 
against this imperative. The Strategy Group should make this one of their prime agenda points.

One more issue requires a novel approach: while the physics community may agree on this exciting 
research agenda, it will ultimately have to convince other scientific disciplines, society and governments. This will be a challenging 
task, more difficult than the case for the LHC, where the community had the ``killer app" of the 
Higgs discovery. Success in this formidable task may  be achieved only if the particle physics 
community at large shows overwhelming support for the recommended programme. Additionally, most of
us need to learn to communicate much more convincingly the prospective
economical, societal, environmental and cultural impacts of HEP research, reaching a wider audience and using modern 
communication means in innovative and exemplary fashion. Professional communication to society will 
become a new required competence. The Strategy Group should also emphasise this. 

\section{Concluding comments}
Exploring in full detail the uniqueness of the Higgs sector and extending substantially the reach 
for New Physics are the tasks at the high-energy frontier for the 21st century. For this agenda, the
FCC-ee/FCC-hh strategy is presently the most promising way forward within and beyond the SM. While 
this programme will arguably revolutionise our physics understanding, its scope implies a new organizational 
mode for CERN, innovative and inspiring ways of collaborating with the next generation of physicists
and more profound and dedicated interaction of the physics community with society. The FCC programme
that we support will require a deep and lasting commitment by society to fundamental research, which the
high-energy community must strive to merit and justify.

\section*{Acknowledgements}
We wish to thank Alain Blondel, Michael Benedikt, Patrick Janot and Frank Zimmermann for their enlightening comments on this document.


%
%
%
%

\end{document}